\documentstyle[prl,aps,epsfig,multicol,amssymb,amsfonts]{revtex}
\tighten
\renewcommand{\narrowtext} 
{\begin{multicols}{2}\global\columnwidth20.5pc} 
\renewcommand{\widetext}
{\end{multicols}\global\columnwidth42.5pc} 
\newcommand{\be}{\begin{equation}}
\newcommand{\ee}{\end{equation}}
\newcommand{\bea}{\begin{eqnarray}}
\newcommand{\eea}{\end{eqnarray}}
\newcommand{\br}{{\bf r}}

\begin{document} 
\draft 
\title{Multifractality at the spin quantum Hall transition} 
\author{F.~Evers$^1$, A.~Mildenberger$^1$,  and A.~D.~Mirlin$^{1,2,*}$ } 
\address{$^1$Institut
f\"ur Nanotechnologie, Forschungszentrum Karlsruhe, 76021 Karlsruhe,
Germany}
\address{$^2$Institut f\"ur Theorie der Kondensierten Materie,
Universit\"at Karlsruhe, 76128 Karlsruhe, Germany}
\date{\today}
\maketitle
\begin{abstract}

Statistical properties of critical wave functions 
at the spin quantum Hall transition are studied
both numerically and analytically (via mapping onto the
classical percolation). It is shown that the index $\eta$
characterizing the decay of wave function correlations is equal to
1/4, at variance with the $r^{-1/2}$ decay of the diffusion propagator. 
The multifractality spectra of eigenfunctions  and of two-point
conductances are found to be close-to-parabolic, $\Delta_q\simeq
q(1-q)/8$ and $X_q\simeq q(3-q)/4$. 

\end{abstract}

\pacs{PACS numbers: 73.43.Cd, 73.43.Nq, 05.45.Df, 05.40.-a} 
\narrowtext

Disordered two-dimensional electron systems show remarkably rich
physics, which is governed by quantum interference effects and
depends on the symmetry class to which the system belongs.
Recently, unconventional symmetry classes
\cite{altland97}, which can be realized in $d$-wave superconductors, have 
attracted considerable interest. Particular attention was paid to
the class C (broken time-reversal invariance), where a transition
between localized phases with quantized values of the spin Hall
conductance takes place \cite{kagalovsky99,senthil99,gruzberg99}.  
A network model describing this spin quantum Hall (SQH) transition was
constructed in \cite{kagalovsky99}, and critical exponents for the scaling
of the localization length were determined numerically. In
\cite{senthil99} a mapping onto a supersymmetric spin chain was
performed, providing an alternative method for the numerical study of the
critical behavior. Remarkably, some exact analytical results for this
model have been obtained by mapping onto the classical percolation
problem \cite{gruzberg99,beamond02}. In particular, it was found that
the density of states (DOS) is critical and scales as
$\rho(\epsilon)\sim \epsilon^{1/7}$,
while the diffusion propagator 
$\Pi({\bf r},{\bf r}')=
\langle G_R({\bf r},{\bf r}')G_A({\bf r}',{\bf r})\rangle$ and
the average two-point conductance $\langle g({\bf r},{\bf r}')\rangle$
fall off as $|{\bf r}-{\bf r}'|^{-1/2}$ at criticality.

The aim of the present paper is to study the statistical
properties of wave functions at the SQH critical point.
Multifractality of wave functions $\psi({\bf r})$ is known to
be a hallmark of the localization transition.
It has been extensively studied in the context of conventional Anderson and
quantum Hall (QH) transitions with {\it non-critical} DOS (see {\it
e.g.} \cite{multifrac} and references therein), and we remind the
reader of some basic results. Multifractality is characterized by a set
of exponents $\tau_q\equiv d(q-1)+\Delta_q$ ($d$ is the spatial
dimensionality) describing the scaling of the
moments of $|\psi^2({\bf r})|$ with the system size, 
$\langle|\psi({\bf r})|^{2q}\rangle \propto L^{-d-\tau_q}$.  
Anomalous dimensions $\Delta_q$ distinguish a
critical point from the metallic phase and determine the scale dependence
of wave function correlations. Among them, $\Delta_2\equiv -\eta$ plays
the most prominent role, governing the spatial correlations of
the ``intensity'' $|\psi|^2$,
\be
L^{2d} \langle |\psi^2({\bf r})\psi^2({\bf r}')|\rangle 
\sim (|{\bf r} - {\bf r}'|/L)^{-\eta}.
\label{e1}
\ee
Correlations of two different (but close in energy) eigenfunctions and
the diffusion propagator possess the same scaling properties,
\bea
&&  L^{2d} \langle |\psi_i^2({\bf r})\psi_j^2({\bf r}')|\rangle,\ 
L^{2d} \langle \psi_i({\bf r})\psi_j^*({\bf r})
\psi_i^*({\bf r}')\psi_j({\bf r}')\rangle, \nonumber \\ 
&& \rho^{-2}\Pi({\bf r},{\bf r}';\omega) 
\sim (|{\bf r} - {\bf r}'|/L_\omega)^{-\eta},
\label{e2}
\eea
where $\omega=\epsilon_i-\epsilon_j$, 
$L_\omega\sim (\rho\omega)^{-1/d}$ and $|{\bf r} - {\bf r}'| < L_\omega$. 
The moments of the two-point conductance show a power-law scaling as
well \cite{janssen99,zirnbauer99}, $\langle
g^q(\br,\br')\rangle\propto|\br-\br'|^{-X_q}$ with another set of
exponents $X_q$, which are related to $\Delta_q$
\cite{klesse01,evers01},    
\be
\label{e3}
X_q=\left\{ \begin{array}{ll}
\Delta_q+\Delta_{1-q}\ , & \qquad q<1/2 \\
2\Delta_{1/2}        \ , & \qquad q>1/2 
\end{array} \right.
\ee       
In two dimensions the multifractal spectra $\Delta_q$ and $X_q$ play a
key role in the identification of the conformal field theory governing
the critical point, which led to growing interest in the
eigenfunction statistics at the QH transition
\cite{janssen99,zirnbauer99,klesse01,evers01,bhaseen00}. 

Applying naively these results to the SQH transition, one would
conclude that the $r^{-1/2}$ scaling of the diffusion propagator found
in \cite{gruzberg99} implies $\eta=1/2$. However, a more careful
examination of the findings of Ref.~\cite{gruzberg99} gives us a
warning: Gruzberg {\it et al.} find that $\Pi$ and $\langle g\rangle$
share the same  $r^{-1/2}$ scaling, in contrast to Eqs.~(\ref{e2}),
(\ref{e3}) predicting different exponents ($\eta\equiv-\Delta_2$ and
$2\Delta_{1/2}$, respectively). This suggests that one should be 
cautious when trying to apply the above relations between critical
exponents obtained for systems with a non-critical DOS to those with a
critical one (like the SQH transition). Our analysis below shows that
such a caution is well justified. 

\begin{figure}
\centerline{
\includegraphics[width=0.9\columnwidth,clip]{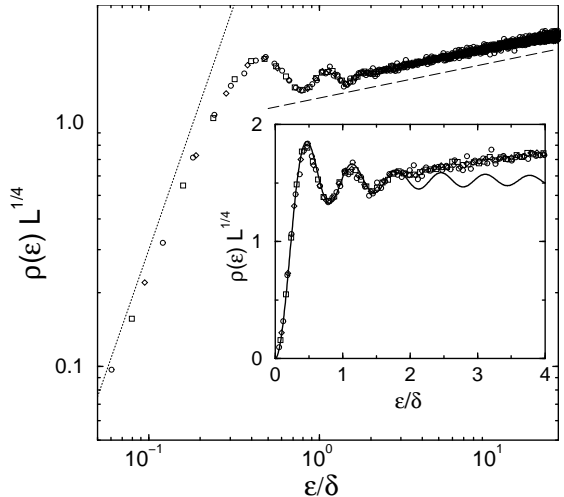}}
\vspace{3mm}
\caption{Scaling plot of the density of states for system sizes
$L=16 (\diamond), 32 (\Box), 96 (\circ)$.
Dashed and dotted lines indicate power laws
(dashed: $\epsilon^{1/7}$, dotted: $\epsilon^{2}$),
$\delta=1/2\pi L^{7/4}$ denotes the level spacing.
Inset: same data on a linear scale and the result from
random matrix theory \protect\cite{altland97} (solid curve). 
}
\label{fig1}
\end{figure}

After this introduction, we turn to our study, which combines
numerical and analytical methods. Our main aim is to
calculate exponents governing the scaling of $\langle
|\psi|^{2q}\rangle$, $\langle g^q\rangle$, and $\Pi$, and to understand
relations between them. For computer simulations we used the SU(2)
network model described in \cite{kagalovsky99,beamond02}.
Each realization of the network is characterized by a unitary
evolution operator $\cal U$. Diagonalizing $\cal U$ using advanced
sparse matrix packages \cite{numerics} yields
eigenfunctions $\psi_{i}$ and eigenvalues $e^{-i\epsilon_i}$,
where $i=1,2,\ldots L^2$. In Fig.~\ref{fig1}
we display the DOS for different system sizes $L$. It is seen that after a
proper rescaling all data collapse onto a single curve, as expected at
criticality. At $\epsilon\gg\delta$ (where $\delta\sim L^{-7/4}$ is
the level spacing at $\epsilon=0$) the DOS scales as
$\rho(\epsilon)\sim \epsilon^{1/7}$, in agreement with the analytical
prediction \cite{gruzberg99}. On the other hand, at
$\epsilon\sim\delta$ one observes an oscillatory structure
qualitatively analogous to the behavior found in the random matrix
theory for the class C \cite{altland97}.

\begin{figure}
\centerline{
\includegraphics[width=0.9\columnwidth,clip]{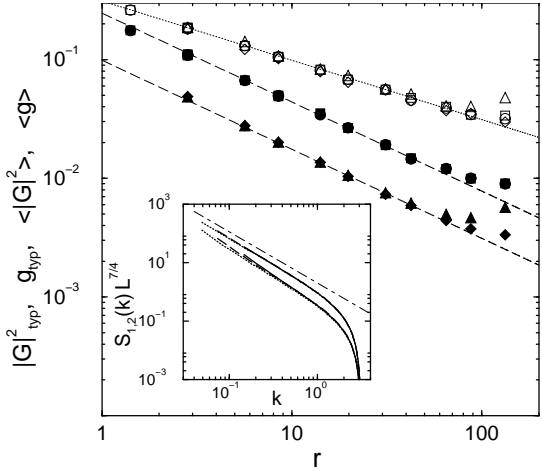}}
\vspace{3mm}
\caption{Scaling of the two-point conductance
with distance $r$ between the contacts: 
average value (empty symbols), $\langle g\rangle$,
and typical value (filled symbols),
$g_{{\rm typ}}=\exp \langle \ln g\rangle$,
in systems of sizes $L=128 (\Box)$ and $L=196 (\circ)$.
Also shown is scaling of the two-point Green function, $\langle
|G|^2 \rangle$ and 
$|G|^2_{{\rm typ}}=\exp \langle \ln |G|^2 \rangle$ 
($L=128 (\bigtriangleup), L=196 (\diamond)$). The
lines correspond to the $r^{-1/2}$ (dotted) and $r^{-3/4}$ (dashed)
power laws. Deviations from power-law
scaling at large values of $r$ are due to the finite system size.
Inset: Fourier transforms of the one- and two-eigenfunction
correlation functions, $S_1(r)=\langle
|\psi^2_{i\alpha}(e)\psi^2_{i\beta}(e')|\rangle$ (upper curves) and
$S_2(r)=\langle\psi^*_{i\alpha}(e)\psi_{j\alpha}(e)
\psi_{i\beta}(e')\psi^*_{j\beta}(e')\rangle$ (lower curves) for
$\epsilon_{i,j}\sim \delta$ and $L=128$ (solid), $256$ (dashed),
$384$ (dotted). The dot-dashed line indicates a
power law $S(k) \propto k^{-7/4}$ corresponding to $S(r)\propto
r^{-1/4}$.} 
\label{fig2}
\end{figure}

We concentrate now on the
statistics of eigenfunctions with smallest energies,
$|\epsilon|\sim\delta$, for which the correlation length
$\xi_\epsilon\sim\epsilon^{-4/7}$ \cite{gruzberg99} is of the order of
the system size. In Fig.~\ref{fig2} (inset) we plot the eigenfunction
autocorrelation function 
(\ref{e1}), as well as the correlation function 
of two eigenstates neighboring in energy (the
second one in Eq.~(\ref{e2})).
The results imply a power-law behavior $\propto r^{-\eta}$ of both
correlation functions at distances $1\ll r\ll L$, with an index
$\eta$ close to $1/4$. On the other hand, our numerics confirms the
value 1/2 \cite{gruzberg99} of the exponent governing the decay of
$\Pi$ (Fig.~\ref{fig2}, main panel).  To understand the difference between the
two exponents, we turn now to an analytical approach. 

Consider a correlation function of two wavefunctions,
\bea
\label{e4}
{\cal D}(e',e;\epsilon_1,\epsilon_2)&=&\langle\sum_{ij\alpha\beta}
\psi_{i\alpha}^*(e)\psi_{j\alpha}(e)\psi_{i\beta}(e')\psi_{j\beta}^*(e')
\nonumber \\
&\times & \delta(\epsilon_1-\epsilon_i)\delta(\epsilon_2-\epsilon_j)\rangle,
\eea
where $e$, $e'$ are two edges of the network and 
$\alpha,\beta=1,2$ are the SU(2) indices.
Introducing the Green function
$G(e',e;z)=\langle e'|(1-z{\cal U})^{-1}|e\rangle$, we express
(\ref{e4}) as
\bea
\label{e5}
&& {\cal D}(e',e;\epsilon_1,\epsilon_2)=(2\pi)^{-2}
\langle{\rm Tr}[G_R(e',e;e^{i\epsilon_1})-G_A(e',e;e^{i\epsilon_1})]
\nonumber\\ 
&& \qquad
\times[G_R(e,e';e^{i\epsilon_2})-G_A(e,e';e^{i\epsilon_2})]\rangle, 
\eea
where $G_{R,A}$ are retarded and advanced Green functions,
$G_{R,A}(e',e;e^{i\epsilon_1})=G(e',e;e^{i(\epsilon_1\pm i0)})$.
We will calculate (\ref{e5}) at zero energy, $\epsilon_{1,2}\to 0$, but
finite level broadening, $\pm i0\to \pm i\gamma$. The scaling behavior
of the correlation function (\ref{e4}) at
$\epsilon_1,\epsilon_2\sim\epsilon$ can then be obtained by
substituting $\epsilon$ for $\gamma$. We thus need to calculate
\bea
\label{e6}
{\cal D}(e',e;\gamma)&=&(2\pi)^{-2}
\langle{\rm Tr}[G(e',e;z)-G(e',e;z^{-1})] \nonumber \\
&\times & [G(e,e';z)-G(e,e';z^{-1})]\rangle,
\eea
with a real $z=e^{-\gamma}<1$. To do this, we make use of the mapping
to the classical percolation, following the approach of
\cite{beamond02}. We give only a brief outline of the calculation
here; details will be published elsewhere \cite{in_preparation}.
The Green functions in (\ref{e6}) are represented as sums over
paths; the resulting expression is to be averaged over SU(2)
matrices $U_f$ associated with all  network edges $f$. The crucial
point is that for each edge $f$ only paths visiting it 0 or 2 times are to
be taken into account. In \cite{beamond02} this was proven for the
average Green function $\langle{\rm Tr}G(e,e;z)\rangle$. The proof
is based on the observation that 
$\langle U_f^q\rangle =c_q\cdot {\bf 1}$, where $c_q=0$ for integer $q\ne
0,\pm 2$. We generalize the statement to products of two Green
functions of the type entering (\ref{e6}) as well as another two-point
correlation function,
\bea
\label{e7}
\tilde{D}(e',e;\gamma)&=&(2\pi)^{-2}
\langle{\rm Tr}[G(e,e;z)-G(e,e;z^{-1})] \nonumber\\
&\times & [G(e',e';z)-G(e',e';z^{-1})]\rangle,
\eea
corresponding to the $\langle|\psi_i^2(e)\psi_j^2(e')|\rangle$
correlator, in the following way. Classifying the paths according to
the number of times they return to a link $f$, we obtain expressions
of the type $\sum_{q_1,q_2=1}^\infty\langle {\rm
Tr}U_f^{q_1}AU_f^{q_2}B\rangle x^{q_1+q_2}$ with $A,B\in {\rm SU(2)}$ and
$x\in\mathbb{R}$. Averaging over $U_f$ yields now two terms,
proportional to $c_{q_1+q_2}$  and $c_{q_1-q_2}$, respectively. While
the first one is non-zero only for $q_1+q_2=2$ traversals of the link
as required, the second one seems to spoil the proof. However,
summing over $q_1$ at fixed $q_1+q_2$, we find that such terms cancel
in view of $\sum_q c_q =c_{-2}+c_0+c_2=-1/2+1+(-1/2)=0$.
Having established that only paths visiting each link 0 or 2 times are to be
considered, we can express, in analogy with \cite{beamond02},
the correlation functions (\ref{e6}) and
(\ref{e7}) in terms of sums over paths representing hulls in
the percolation problem. In particular, we get for the
products of Green functions entering (\ref{e6}) 
\bea
\langle{\rm Tr} G(e',e;z)G(e,e';z)\rangle & = &
\langle{\rm Tr} G(e',e;z^{-1})G(e,e';z^{-1})\rangle \nonumber \\ 
&=& -2\sum _N P(e',e;N)z^{2N}, \label{e8} \\
\langle{\rm Tr} G(e',e;z) G(e,e';z^{-1})\rangle &=&
-2\sum_N P_1(e',e;N)z^{2N}, \label{e9} 
\eea
where $P(e',e;N)$ and $P_1(e',e;N)$ are probabilities that the edges
$e$ and $e'$ belong to the same loop of the length $N$ (resp. with the
length $N$ of the part corresponding to the motion from $e$ to
$e'$). At $z=1$ both (\ref{e8}) and (\ref{e9}) reduce to $-2$ times
the probability $P(e',e)$ that $e$ and $e'$ belong to the same loop,
which is identical (up to the sign) to the expression for the average
conductance $\langle g(e',e)\rangle$ at $\epsilon=0$ obtained in
\cite{beamond02}. 

The fractal dimension of the percolation
hulls is 7/4 \cite{saleur87}, implying \cite{moore02} 
that $P$ and $P_1$ scale as
\be
P(r,N),\ P_1(r,N) \sim N^{-8/7}r^{-1/4}\ , \qquad r\lesssim N^{4/7}
\label{e10}
\ee
and fall off exponentially fast at $r\gg N^{4/7}$ ($r$ is the distance
between $e$ and $e'$). This yields for the correlation functions in
(\ref{e8}) and (\ref{e9}) (which we abbreviate as 
$\langle G_R G_R\rangle$, $\langle G_A G_A\rangle$, 
$\langle G_R G_A\rangle$) 
\bea
&& \langle G_R G_R\rangle = \langle G_A G_A\rangle \simeq
\langle G_R G_A\rangle \sim r^{-1/2}, \nonumber\\
&& \hspace*{3cm} r\ll \xi_\gamma\equiv 
\gamma^{-4/7}
\label{e12}
\eea
in full agreement with the scaling argument of
\cite{gruzberg99} and with our numerics. 
However, when we substitute (\ref{e8}), (\ref{e9})
in (\ref{e6}), these leading order terms cancel since
$\sum_N[P(r,N)-P_1(r,N)]=0$. The result is
non-zero due to the factors $z^{2N}$ only, implying that relevant $N$
are now $N\sim \gamma^{-1}$, so that $\langle(G_R-G_A)(G_R-G_A)\rangle$
scales differently compared to (\ref{e12}),
\bea
{\cal D}(e',e;\gamma) & = & {1\over\pi^2}\sum_N
[P(r,N)-P_1(r,N)](1-e^{-2N\gamma}) \nonumber\\
&\sim & P(r,\gamma^{-1})\gamma^{-1}\sim
(\xi_\gamma r)^{-1/4},\qquad r\lesssim \xi_\gamma.
\label{e13}
\eea
Using now the definition (\ref{e4}) of ${\cal D}$ and the DOS scaling,
$\rho(\epsilon)\sim \epsilon^{1/7}\sim \xi_\epsilon^{-1/4}$, we find
for $r\lesssim \xi_\epsilon$
\be
\label{e14}
L^4\langle
\psi_{i\alpha}^*(e)\psi_{j\alpha}(e)\psi_{i\beta}(e')\psi_{j\beta}^*(e')
\rangle
\sim (r/\xi_\epsilon)^{-1/4}.
\ee
The same scaling behavior is obtained for the correlation function 
$\langle|\psi_{i\alpha}^2(e)\psi_{j\beta}^2(e')|\rangle$ 
\cite{in_preparation}.  We thus conclude that
$\eta=1/4$, consistent with our above numerical results. 
To shed more light on the difference in scaling between $\langle
GG\rangle$ (or $\langle g\rangle$), Eq.~(\ref{e12}), and ${\cal D}$,
Eq.~(\ref{e13}), it is instructive to reverse the logic and to ask how
(\ref{e12}) can be obtained from the wave function correlations
(\ref{e13}), (\ref{e14}). It is straightforward to express $\langle
GG\rangle$ through ${\cal D}$ in the form of an integral over
$\epsilon_{1,2}$ with corresponding energy denominators (dispersion
relation). The integral is then dominated by $\epsilon_{1,2}\sim
\epsilon(r)$, where $\epsilon(r)$ is defined by 
$\xi_{\epsilon(r)}\sim r$ ({\it i.e.} $\epsilon(r)\sim r^{-7/4}$).  
This yields 
$\langle GG\rangle \sim {\cal D}(r;\epsilon(r))\sim r^{-1/2}$ 
(we have used (\ref{e13}) in the last step), in agreement with
(\ref{e12}). Therefore, $\langle GG\rangle$ (or $\langle g\rangle$) is
determined by wave functions with energies $\epsilon(r)$, which
transforms $\xi_\epsilon^{-1/4}$ in (\ref{e13}) into an additional
factor $r^{-1/4}$. We will come back to this argument below to get an
analogous relation for higher moments.

\begin{figure}
\centerline{
\includegraphics[width=0.9\columnwidth,clip]{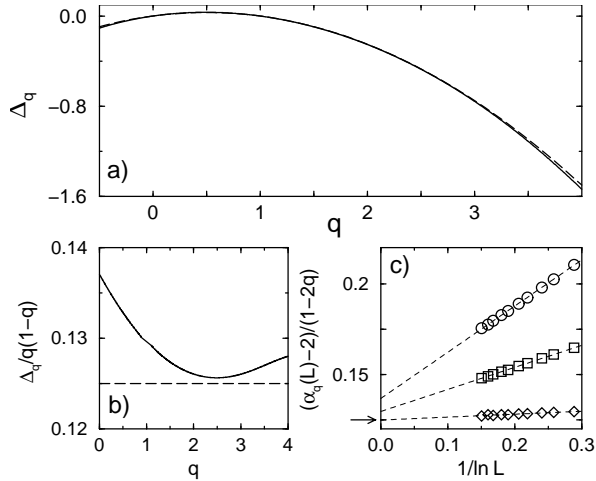}}
\vspace{3mm}
\caption{a) anomalous dimension $\Delta_q$ extrapolated from ensembles with
system sizes $L=16-384$. The dashed line (almost indistinguishable) is
the parabola $\Delta_q=q(1-q)/8$;
b) $\Delta_q/q(1-q)$ over $q$ highlighting weak non-parabolicity;  
c) multifractal exponents
$\alpha_q(L)=-\langle
|\psi|^{2q}\ln|\psi|^2\rangle/\langle|\psi|^{2q}\rangle \ln L $ with
$q=0$ ($\circ$), 1 ($\Box$), and 2 ($\diamond$), and
extrapolation to infinite system size (dashed lines):
$\alpha_0-2=0.137\pm0.003$, $2-\alpha_1=0.130\pm0.003$, 
$(2-\alpha_2)/3=0.125\pm0.001$. 
The arrow indicates the value $\eta/2=1/8$ of these quantities for a
parabolic spectrum. }
\label{fig3}
\end{figure}

In order to study the whole multifractal spectrum $\Delta_q$, we
return to numerical simulations. Our procedure based on the evaluation of
ensemble averaged moments $\langle|\psi^{2q}|\rangle$ and extrapolation
to $L\to\infty$ was described in detail in \cite{evers01}. 
The results for $\tau_q$ are shown in Fig.~\ref{fig3}a. The obtained
spectrum is parabolic with a high accuracy. A parabolic spectrum is
uniquely determined by $\eta$, $\Delta_q=\eta q(1-q)/2$; the
above result $\eta=1/4$ thus implies
\be
\label{e15}
\Delta_q \simeq q(1-q)/8.
\ee
We find, however, clear deviations from the parabolic law (\ref{e15}),
as shown in Fig.~\ref{fig3}b. One could ask whether these are not an
artifact of uncontrollable finite-size corrections to scaling. We
observe, however, an almost perfect scaling for all the quantities
studied, yielding, in particular, $\eta=0.252\pm 0.002$ in a very good
agreement with the exact value $\eta=1/4$. On the other hand, we find
(in standard notations, $\alpha_q=d\tau_q/dq$)
$\alpha_0-2=0.137\pm0.003$ and $2-\alpha_1=0.130\pm0.003$
(Fig.~\ref{fig3}c), while both 
these quantities would be equal to $\eta/2$, should the parabolic law
(\ref{e15}) be exact. This makes us believe that the spectrum is
only approximately parabolic, in contrast to exact parabolicity found
for the QH transition \cite{evers01}.

Finally, we turn to the statistics of two-point
conductances. Generalizing the above argument, we get
\begin{eqnarray*}
&& \langle[{\rm Tr}G(e',e)G(e,e')]^q\rangle \\
&& \sim \rho^{2q}(\epsilon(r))
L^{4q} \langle|\psi^{2q}_{\epsilon_1}(e)\psi^{2q}_{\epsilon_2}(e')|\rangle
|_{\epsilon_{1,2}\sim\epsilon(r)}\sim r^{-X_q},
\end{eqnarray*}
with an index $X_q$ related to $\Delta_q$ and to the scaling dimension
$x_\rho$ of the DOS (defined by
$\rho(\epsilon)\sim\xi_\epsilon^{-x_\rho}$) as follows
(see also \cite{bernard01b}),
\be
X_q = 2qx_{\rho}+2\Delta_q.
\label{e16}
\ee
Using $x_\rho=1/4$ and Eq.~(\ref{e15}) for $\Delta_q$, we find
\be
\label{e17}
X_q \simeq q(3-q)/4.
\ee
The same scaling is expected to hold for the two-point conductance,
$\langle g^q\rangle\sim r^{-X_q}$. However, since $g$ is bounded
from above, $g\le 2$, the exponent for $\langle g^q\rangle$ should be
a non-decreasing function of $q$. Therefore, Eq.(\ref{e17}) will hold
only for the moments $\langle g^q\rangle$ with $q\le q_c$, where
$q_c\simeq 3/2$ is the maximum of (\ref{e17}), while for higher $q$
the exponent saturates, $X_{q\ge q_c}=X_{q_c}\simeq 9/16$ (these moments
are determined by the probability to find $g\sim 1$). 
Equation (\ref{e17}) corresponds to a normal
distribution of $\ln g$ (at $r\gg 1$) with the average 
$\langle \ln g(r)\rangle \simeq -{3\over 4}\ln r$ and the variance 
${\rm var}[\ln g(r)]\simeq {1\over 2}\ln r$. These arguments are
fully confirmed by the results of the numerical simulations shown in
Fig.~\ref{fig2}.

It remains an open question whether the multifractal exponents
$\Delta_q$, $X_q$ can be computed by the conformal field theory
methods 
\cite{zirnbauer99,bhaseen00,bernard01b,read01,bernard01a,fendley00}. 
Note that our results do
not confirm the proposal of Ref.~\cite{bernard01b}, where the value
$\eta=1/2$ was obtained. Apparently, this indicates that the theory
considered in \cite{bernard01b} and obtained \cite{bernard01a} from a
particular network model with fine-tuned couplings, does not belong
to the SQH universality class. 

To summarize, we have studied, by combining numerical and
analytical methods, the wave function statistics at the SQH
transition. In particular, we have shown, using a mapping to 
classical percolation, that the index $\eta\equiv-\Delta_2$ [defined by
Eq.~(\ref{e1})] is equal to 1/4, at variance with the $r^{-1/2}$
scaling of the diffusion propagator $\Pi=\langle G_R G_A\rangle$.
The multifractal spectra of wave functions
($\Delta_q$) and two-point conductances ($X_q$)
are given with a good accuracy by Eqs.~(\ref{e15}) and (\ref{e17}) but
show detectable deviations from parabolicity.

Discussions and correspondence with J.T.~Chalker, I.~Gruzberg,
A. LeClair, and J.E.~Moore are gratefully acknowledged. 
This work was supported by the SFB 195 and the Schwerpunktprogramm 
"Quanten-Hall-Systeme"  der Deutschen Forschungsgemeinschaft.


\end{multicols}
\end{document}